\documentclass[aps,prd,notitlepage,showpacs,nofootinbib,superscriptaddress]{revtex4-1}

\usepackage{graphicx}
\usepackage[utf8]{inputenc} 

\usepackage{amsmath}
\usepackage{amssymb}
\usepackage{float}
\usepackage{comment}
\usepackage{slashed}
\usepackage[normalem]{ulem}

\usepackage{booktabs}
\usepackage{mathtools,braket,blkarray}

\usepackage[usenames,dvipsnames]{color}
\usepackage[colorinlistoftodos]{todonotes}
\usepackage[colorlinks=true,citecolor=darkred,urlcolor=darkred, pdfborder={0 0 0}]{hyperref}
\usepackage[normalem]{ulem}

\usepackage{multirow}
\definecolor{darkred}{rgb}{0.6,0,0}
\usepackage[colorinlistoftodos]{todonotes}

\definecolor{linkcolor}{rgb}{0,0,0.5}
 \newcommand {\ignore}[1]{}

\def \znbb {$\rm 0\nu\beta\beta$ }

\def\gsim{\raise0.3ex\hbox{$\;>$\kern-0.75em\raise-1.1ex\hbox{$\sim\;$}}}
\def\lsim{\raise0.3ex\hbox{$\;<$\kern-0.75em\raise-1.1ex\hbox{$\sim\;$}}}

\newcommand{\sm}{{Standard Model }}

\usepackage{soul}

\definecolor{mightnightblue}{RGB}{25,25,112}

\definecolor{brown}{rgb}{0.59, 0.29, 0.0}

\def\vev#1{\left\langle #1\right\rangle}

\def\21{$\mathrm{SU(2)_L \otimes U(1)_Y}$}

\def\sm{standard model }

\newcommand{\AddrAHEP}{%
  AHEP Group, Institut de F\'{i}sica Corpuscular --
  C.S.I.C./Universitat de Val\`{e}ncia, Parc Cient\'ific de Paterna.\\
  C/ Catedr\'atico Jos\'e Beltr\'an, 2 E-46980 Paterna (Valencia) - SPAIN}

\newcommand{\AddrUNAM}{ {\it Instituto de F\'{\i}sica, Universidad Nacional Aut\'onoma de M\'exico, A.P. 20-364, Ciudad de M\'exico 01000, M\'exico.}}

\begin{document}

\bibliographystyle{unsrt}

\title{\boldmath \color{BrickRed} Probing the predictions of an orbifold theory of flavor}

\author{Francisco J. de Anda}\email{fran@tepaits.mx}
\affiliation{Tepatitl{\'a}n's Institute for Theoretical Studies, C.P. 47600, Jalisco, M{\'e}xico}

\author{Newton Nath}\email{newton@fisica.unam.mx}
\affiliation{\AddrUNAM}

\author{Jos\'{e} W. F. Valle}\email{valle@ific.uv.es}
\affiliation{\AddrAHEP}

\author{Carlos A. Vaquera-Araujo}\email{vaquera@fisica.ugto.mx}
\affiliation{Consejo Nacional de Ciencia y Tecnolog\'ia, Av. Insurgentes Sur 1582. Colonia Cr\'edito Constructor, Del. Benito Ju\'arez, C.P. 03940, Ciudad de M\'exico, M\'exico}
\affiliation{Departamento de F\'isica, DCI, Campus Le\'on, Universidad de
  Guanajuato, Loma del Bosque 103, Lomas del Campestre C.P. 37150, Le\'on, Guanajuato, M\'exico}

%\preprint{IPM/P-2012/009}
%\vspace*{3mm}

\begin{abstract}
\vspace{0.5cm}

We examine the implications of a recently proposed theory of fermion masses and mixings in which an $A_4$ family symmetry emerges from orbifold compactification.
We analyse two variant schemes concerning their predictions for neutrino oscillations, neutrinoless double-beta decay and the golden quark-lepton unification mass relation.
We find that upcoming experiments DUNE as well as LEGEND and nEXO offer good chances of exploring a substantial region of neutrino parameters.
  
\end{abstract}
%%%%%%%%%%%%

\maketitle
\noindent

%%%%%%%%%%%%%%%%%%%%%%%%%%%%%%%%%%%%
\section{Introduction}
%%%%%%%%%%%%%%%%%%%%%%%%%%%%%%%%%%%%

The discovery of neutrino oscillations~\cite{Kajita:2016cak,McDonald:2016ixn} has prompted a great experimental effort toward precision measurements~\cite{deSalas:2017kay}. 
Indeed, the pattern of neutrino mass and mixing parameters is  strikingly at odds with the one that characterizes the quark sector, suggesting that it can hardly be expected to happen just by chance.
The most popular approach to bring a \textit{rationale} to the pattern of neutrino mixing involves the idea that there is some non-Abelian family symmetry in nature.
 %%%
In a model-independent way one may assume the existence of some residual CP symmetry characterizing the neutrino mass matrix, irrespective of the details of the underlying
theory~\cite{Chen:2015siy,Chen:2016ica,Chen:2019fgb}.
A more ambitious approach is, of course, to guess what the family symmetry actually is, and to build explicit flavor models on a case-by-case basis~\cite{babu:2002dz,Altarelli:2010gt,Morisi:2012fg,King:2014nza,Chen:2015jta}. However, pinning down the nature of such symmetry among the plethora of possibilities is a formidable task.

An interesting theoretical idea has been to imagine the existence of new dimensions in space-time, as a way to shed light on the possible nature of the family symmetry in four dimensions.
In this context, six-dimensional theories compactified on a torus have been suggested~\cite{deAnda:2018oik,deAnda:2018yfp} and a realistic \sm extension has recently been
proposed~\cite{deAnda:2019jxw} in which fermions are nicely arranged within the framework of an $A_4$ family symmetry.
The theory yields very good predictions for fermion masses and mixings, including the ``golden'' quark-lepton unification formula~\cite{Morisi:2011pt,King:2013hj,Morisi:2013eca,Bonilla:2014xla,Reig:2018ocz}.

In this work we focus on the possibility of probing the implications of this theory within the next generation of neutrino experiments.
This includes the long-baseline oscillation experiment DUNE~\cite{Acciarri:2015uup, Alion:2016uaj} as well as neutrinoless double-beta decay (\znbb for short) searches.
In Sec.~\ref{sec:TheFrame} we describe the theory framework, identifying two model setups, while in Sec.~\ref{sec:neutrino-predictions} we determine the potential of upcoming neutrino experiments, such as DUNE and \znbb experiments to probe our orbifold compactification predictions.

%%%%%%%%%%%%%%%%%%%%%%%%%%%%%%%%%%%%
\section{Theory Framework}
\label{sec:TheFrame}

Our model features a 6-dimensional version of the Standard Model $SU(3)\otimes SU(2)\otimes U(1)$ gauge symmetry, together with 3 right handed neutrinos and supplemented with the orbifold compactification described in our previous paper \cite{deAnda:2019jxw}.
The transformation properties of the fields under the gauge and $A_4$ family symmtetry and their localization on the orbifold are shown in table \ref{tab:fields}.

\begin{table}[h]
\centering
\footnotesize 
%\captionsetup{width=0.9\textwidth}
\begin{tabular}{ |cc|ccc|cccccc|}
\hline
\textbf{Field} &$\qquad$ & $SU(3)$ & $SU(2)$ & $U(1)$ &$\qquad$  &$A_4$ &$\qquad$& $\mathbb{Z}_3$& $\qquad$& Localization \\
\hline
$L$ & & $\mathbf{1}$ & $\mathbf{2}$ & $-1/2$ & & $\mathbf{3}$  & & $\omega^2$&  &Brane\\
$d^c$ && $\bar{\mathbf{3}}$ & $\mathbf{1}$ & $1/3$ & & $\mathbf{3}$  & & $\omega$&& Brane\\
$e^c$ &&$\mathbf{1}$ & $\mathbf{1}$ & $1$ & & $\mathbf{3}$  & & $\omega$&& Brane\\
$Q$ && $\mathbf{3}$ & $\mathbf{2}$ & $1/6$ & & $\mathbf{3}$  & & $\omega^2$&&Brane\\
$u_{1,2,3}^c$ && $\bar{\mathbf{3}}$ & $\mathbf{1}$ & $-2/3$ & & $\mathbf{1''},\mathbf{1'},\mathbf{1}$ & & $\omega^2$&&Bulk \\
$\nu^c$ && $\mathbf{1}$ & $\mathbf{1}$ & $0$ & & $\mathbf{3}$   & & $1$&&Brane \\
\hline
$H_u$ && $\mathbf{1}$ & $\mathbf{2}$ & $1/2$ & & $\mathbf{3}$ & & $\omega^2$&& Brane \\
$H_d$  && $\mathbf{1}$ & $\mathbf{2}$ & $-1/2$ & & $\mathbf{3}$  & & $1$&&Brane\\
$H_\nu$ && $\mathbf{1}$ & $\mathbf{2}$ & $1/2$ & & $\mathbf{3}$ & & $\omega$&& Brane \\
$\sigma$ && $\mathbf{1}$ & $\mathbf{1}$ & $0$ & & $\mathbf{3}$ & & $1$&& Bulk\\
\hline
\end{tabular}
\caption{Field content of the model.}
\label{tab:fields}
\end{table}

The scalar sector consists of three Higgs doublets and an extra singlet scalar $\sigma$, all transforming as flavor triplets.  They are charged under a $\mathbb{Z}_3$ symmetry, so that $H_d$ only couples to down-type fermions (charged leptons and down quarks), $H_u$ couples only to up-quarks and $H_\nu$ only couples to neutrinos.

The effective Yukawa terms are given by
\begin{equation}
\begin{split}
\mathcal{L}_Y &= y^N  \nu^c \nu^c \sigma+y_1^\nu(L H_\nu \nu^c)_1+y_2^\nu(L H_\nu \nu^c)_2\\
 &\quad +y_1^d(Q d^c H_d)_1+y_2^d(Q d^c H_d)_2+y_1^e(L e^c H_d)_1+y_2^e(L e^c H_d)_2\\
&\quad +y_1^u(QH_u)_{1'}u_{1}^c+y_2^u(QH_u)_{1''}u_{2}^c+y_3^u(QH_u)_{1}u_{3}^c,
\label{eq:yuk}
\end{split}
\end{equation}
where the symbol $()_{1,2}$ indicates the possible singlet contractions $\mathbf{3}\times \mathbf{3} \times \mathbf{3}\to \mathbf{1}_{1,2}$ and $\mathbf{3}\times \mathbf{3} \to \mathbf{1}_{1,1',1''}$ in $A_4$. All dimensionless Yukawa couplings are assumed to be real due to a CP symmetry.

The scalar field $\sigma$ gets a vacuum expected value (VEV) that breaks spontaneously lepton number and the $A_4$ family symmetry, giving large Majorana masses to the right handed neutrinos.
The corresponding VEV is aligned as
\begin{equation}
\braket{\sigma}=v_\sigma\left(\begin{array}{c}1\\ \omega\\ \omega^2\end{array}\right),
\end{equation}
with $\omega=e^{2\pi i/3}$, the cube root of unity.

As $A_4$ is broken at a high mass scale, the Higgs doublets can obtain the most general spontaneous CP violating alignment, which we parametrize as
\begin{equation}\begin{split}
\braket{H_u}=v_u\left(\begin{array}{c}\epsilon_1^u e^{i\phi_1^u}\\ \epsilon_2^u e^{i\phi_2^u}\\ 1\end{array}\right),\ \ \ \ \braket{H_\nu}=v_\nu e^{i\phi^\nu}\left(\begin{array}{c}\epsilon_1^\nu e^{i\phi_1^\nu}\\ \epsilon_2^\nu e^{i\phi_2^\nu}\\ 1\end{array}\right),\ \ \ \braket{H_d}=v_d e^{i\phi^d}\left(\begin{array}{c}\epsilon_1^d e^{i\phi_1^d}\\ \epsilon_2^d e^{i\phi_2^d}\\ 1\end{array}\right).
\end{split}\end{equation}

An important prediction of the model comes from the fact that the charged leptons and down-quarks obtain their masses from the same $H_d$, so that the $A_4$ structure implies the golden relation between their masses \cite{Morisi:2011pt}
\begin{equation}\label{eq:golden}
\frac{m_\tau}{\sqrt{m_\mu m_e}}=\frac{m_b}{\sqrt{m_s m_d}},
\end{equation}
This relation is in good agreement with experiments~\cite{Tanabashi:2018oca} and is rather robust against renormalization group running.

The explicit form of the mass matrices for the matter fields (up to unphysical rephasings) is given as
\begin{equation}\begin{split}
 M_u&=v_u\left(\begin{array}{ccc} y_1^u\epsilon_1^u &y_2^u \epsilon_1^u & y_3^u\epsilon_1^u \\
 y_1^u\epsilon_2^u \omega^2&  y_2^u\epsilon_2^u \omega &   y_3^u\epsilon^u_2 \\
  y_1^u \omega& y_2^u \omega^2&y_3^u\end{array}\right),\\
M_d&=v_d\left(\begin{array}{ccc} 0 & y_1^d\epsilon_1^d e^{i(\phi_1^d-\phi_2^d)}& y_2^d \epsilon_2^d \\
 y_2^d\epsilon_1^d e^{i(\phi_1^d-\phi_2^d)}& 0 &  y_1^d\\
 y_1^d\epsilon_2^d & y_2^d&0\end{array}\right),\\
M_e&=v_d\left(\begin{array}{ccc} 0 & y_1^e\epsilon_1^d e^{-i(\phi_1^d-\phi_2^d)}& y_2^e \epsilon_2^d \\
 y_2^e\epsilon_1^d e^{-i(\phi^1_d-\phi_2^d)}& 0 &  y_1^e\\
 y_1^e\epsilon_2^d & y_2^e&0\end{array}\right),\\
M_N^R&=y^N v_\sigma\left(\begin{array}{ccc}0 &   \omega^2 & \omega \\  \omega^2 & 0&1\\  \omega&1&0\end{array}\right),\\
M_\nu^D&=v_\nu\left(\begin{array}{ccc} 0 & y_1^\nu\epsilon_1^\nu e^{i(\phi^\nu_1-\phi_2^\nu)} & y_2^\nu \epsilon_2^\nu \\
 y_2^\nu\epsilon_1^\nu e^{i(\phi_1^\nu-\phi_2^\nu)}& 0 &  y_1^\nu\\
 y_1^\nu\epsilon_2^\nu & y_2^\nu&0\end{array}\right),\\
M_\nu^L&=M_\nu^D(M_N^R)^{-1}(M_\nu^D)^T.
\label{eq:massmat}
\end{split}\end{equation}

In what follows we adopt the standard parametrization for the Cabibbo-Kobayashi-Maskawa (CKM) matrix
\begin{equation}  \label{eq:CKM}
 V_{CKM}= \left( 
\begin{array}{ccc}
c^q_{12} c^q_{13} & s^q_{12} c^q_{13}  & s^q_{13} e^{-i\delta^q}
\\ 
-s^q_{12} c^q_{23}- c^q_{12} s^q_{13} s^q_{23} e^{ i \delta^q} & c^q_{12} c^q_{23} - s^q_{12} s^q_{13} s^q_{23} e^{ i \delta^q } & c^q_{13} s^q_{23} \\ 
s^q_{12} s^q_{23} - c^q_{12} s^q_{13} c^q_{23} e^{ i\delta^q } & - c^q_{12} s^q_{23}  - s^q_{12} s^q_{13} c^q_{23} e^{i\delta^q } & c^q_{13} c^q_{23}%
\end{array}
\right)\,, 
\end{equation}
and the symmetrical presentation of the lepton mixing matrix~\cite{Schechter:1980gr,Rodejohann:2011vc},  
\begin{equation}  \label{eq:symmetric_para}
 K= \left( 
\begin{array}{ccc}
c^{\ell}_{12} c^{\ell}_{13} & s^{\ell}_{12} c^{\ell}_{13} e^{ - i \phi_{12} } & s^{\ell}_{13} e^{ -i \phi_{13} }
\\ 
-s^{\ell}_{12} c^{\ell}_{23} e^{ i \phi_{12} } - c^{\ell}_{12} s^{\ell}_{13} s^{\ell}_{23} e^{ -i ( \phi_{23} -
\phi_{13} ) } & c^{\ell}_{12} c^{\ell}_{23} - s^{\ell}_{12} s^{\ell}_{13} s^{\ell}_{23} e^{ -i ( \phi_{23} +
\phi_{12} - \phi_{13} ) } & c^{\ell}_{13} s^{\ell}_{23} e^{- i \phi_{23} } \\ 
s^{\ell}_{12} s^{\ell}_{23} e^{ i ( \phi_{23} + \phi_{12} ) } - c^{\ell}_{12} s^{\ell}_{13} c^{\ell}_{23} e^{ i
\phi_{13} } & - c^{\ell}_{12} s^{\ell}_{23} e^{ i \phi_{23} } - s^{\ell}_{12} s^{\ell}_{13} c^{\ell}_{23} e^{
-i ( \phi_{12} - \phi_{13} ) } & c^{\ell}_{13} c^{\ell}_{23}%
\end{array}
\right)\,,
\end{equation}
with $c^{f}_{ij}\equiv \cos\theta^f_{ij}$ and $s^{f}_{ij}\equiv \sin\theta^f_{ij}$ where $f=q,\ell$. The advantage of using the symmetrical parametrization for the lepton mixing matrix resides in the transparent role of the Majorana phases in the effective mass parameter characterizing the amplitude for neutrinoless double beta decay
\begin{equation}
\langle m_{\beta\beta}\rangle=\left|\sum_{j=1}^3 K_{ej}^2 m_j\right|=
\left|c^{\ell\,2}_{12}c^{\ell\,2}_{13} m_1 + s^{\ell\,2}
_{12}c^{\ell\,2}_{13} m_2 e^{2i\phi_{12} }+ s^{\ell\,2}_{13} m_3 e^{2i\phi_{13}}\right|~,
\end{equation}
while keeping a rephasing-invariant expression for the Dirac phase
\begin{equation}
  \delta^{\ell}=\phi_{13}-\phi_{12}-\phi_{23}
  \label{eq:dell}
\end{equation}
which affects neutrino oscillation probabilities. 

%%%%%%%%%%%%%%%%%%%%%%%%%%%%%%%%%%%%%%%%%%%%%%%%%%
\subsection{Model Set-up I (MI)}\label{sec:M1}
%%%%%%%%%%%%%%%%%%%%%%%%%%%%%%%%%%%%%%%%%%%%%%%%%%

Following  \cite{deAnda:2019jxw}, we may further assume that the Higgs's VEVs preserve conventional (trivial) CP symmetry, and therefore they are real.
Together with the reality of the Yukawa couplings, this implies that the only source of CP violation is the phase $\omega$. This leads to a very strong predictivity.

The model is specified by 15 parameters ($y_{1,2}^{\nu}v_\nu,y_{1,2}^{e,d}v_d,\ y_{1,2,3}^uv_u,\ \epsilon_{1,2}^{u,\nu,d}$) that describe 22 low-energy flavor observables:
($m_{u,c,t,d,s,b,e,\mu,\tau},\ m^{\nu}_{1,2,3},\ \theta_{12,13,23}^q,\ \delta^q,\ \theta^l_{12,13,23}, \phi_{12,13,23}$) including the neutrino Majorana phases.
One extra parameter ($y^Nv_\sigma$) defines the masses of the 3 right handed neutrinos.
\begin{table}[ht]
	\centering
	\footnotesize
	\renewcommand{\arraystretch}{1.1}
	\begin{tabular}[t]{|lc|r|}
		\hline
		Parameter &\qquad& Value \\ 
		\hline
		$y^e_1v_d/\mathrm{GeV}$ &\quad& $1.745$ \\
		$y^e_2v_d/(10^{-1}\mathrm{GeV})$ && $-1.019$ \\
		\rule{0pt}{3ex}%
		$y^d_1v_d/(10^{-2}\mathrm{GeV})$ &&$-4.690$ \\
		$y^d_2v_d /\mathrm{GeV}$ && $2.88$ \\
		\rule{0pt}{3ex}%
		$y^\nu_{1} v_\nu/\sqrt{Y^N v_\sigma \mathrm{meV}\times10^{-1}}$ &&$7.54$ \\
		$y^\nu_2v_\nu/(\sqrt{Y^N v_\sigma \mathrm{meV}}\times10^{-3} )$ && $1.89$ \\
		\rule{0pt}{3ex}%
		$y^u_{1} v_u/(10^{-1}\mathrm{GeV})$ &&$6.24$ \\
		$y^u_2v_u/(10^2\mathrm{GeV})$ && $1.71$ \\
		$y^u_3v_u/\mathrm{GeV}$ && $-7.13$ \\
		\rule{0pt}{3ex}%
		$\epsilon^u_1/10^{-4}$ && $-6.90 $ \\
		$\epsilon^u_2/10^{-2}$ && $6.24$ \\
		\rule{0pt}{3ex}%
		$\epsilon^d_1/10^{-3}$ && $-2.74$ \\
		$\epsilon^d_2/10^{-3}$ && $6.00$ \\
		\rule{0pt}{3ex}%
		$\epsilon^\nu_1$ && $1.16$ \\
		$\epsilon^\nu_2/10^{-1}$ && $-3.23$ \\
		\hline	
	\end{tabular}
	\hspace*{0.5cm}
	\begin{tabular}[t]{ |l |c|c c |c| }
		\hline
\multirow{2}{*}{Observable}& \multicolumn{2}{c}{Data} & & \multirow{2}{*}{Model best fit}  \\
		\cline{2-4}
		& Central value & 1$\sigma$ range  &   & \\
		\hline
		$\theta_{12}^\ell$ $/^\circ$ & 34.44 & 33.46 $\to$ 35.67 && $34.36$  \\ 
		$\theta_{13}^\ell$ $/^\circ$ & 8.45 & 8.31 $\to$ 8.61  && $8.31$  \\  
		$\theta_{23}^\ell$ $/^\circ$ & 47.69 & 45.97 $\to$ 48.85  && $48.47$ \\ 
		$\delta^\ell$ $/^\circ$ & 237 & 210 $\to$ 275 && $268$  \\
		$m_e$ $/ \mathrm{MeV}$ & 0.489 &  0.489 $\to$ 0.489 && $0.489$ \\ 
		$m_\mu$ $/  \mathrm{GeV}$ & 0.102 & 0.102  $\to$ 0.102  &&  $0.102$ \\ 
		$m_\tau$ $/ \mathrm{GeV}$ &1.745 & 1.743 $\to$1.747 && $1.745$ \\ 
		$\Delta m_{21}^2 / (10^{-5} \, \mathrm{eV}^2 ) $ & 7.55  & 7.39 $\to$ 7.75 && $7.63$  \\
		$\Delta m_{31}^2 / (10^{-3} \, \mathrm{eV}^2) $ & 2.50  & 2.47 $\to$ 2.53 &&  $2.42$ \\
		$m_1$ $/\mathrm{meV}$  & & & & $4.12$ \\ 
		$m_2$ $/\mathrm{meV}$  & && & $9.66$ \\ 
		$m_3$ $/\mathrm{meV}$  & && & $50.11$ \\
		$ \phi_{12} $ $/^\circ$ & & && $250$  \\
		$ \phi_{13}$  $/^\circ$& & && $187$  \\    
		$ \phi_{23}$  $/^\circ$& & && $29$  \\    	
		\hline
		$\theta_{12}^q$ $/^\circ$ &13.04 & 12.99 $\to$ 13.09 &&  $13.04$ \\	
		$\theta_{13}^q$ $/^\circ$ &0.20 & 0.19 $\to$ 0.22 && $0.20$  \\
		$\theta_{23}^q$ $/^\circ$ &2.38& 2.32 $\to$ 2.44 && $2.37$  \\	
		$\delta^q$ $/^\circ$ & 68.75 & 64.25 $\to$ 73.25  & & $60.25$\\
		$m_u$ $/ \mathrm{MeV}$ & 1.28 & 0.76$\to$ 1.55 && $1.29$  \\	
		$m_c$ $/ \mathrm{GeV}$ & 0.626 & 0.607 $\to$ 0.645 &&  $0.626$ \\	
		$m_t$ $/\mathrm{GeV}$  	  & 171.6& 170 $\to$ 173  && $171.6$ \\
		$m_d$ $/ \mathrm{MeV}$ & 2.74 & 2.57 $\to$ 3.15 &&  $2.75$ \\	
		$m_s$ $/ \mathrm{MeV}$ & 54 & 51 $\to$ 57 && $51$ \\
		$m_b$ $/ \mathrm{GeV}$	  & 2.85 &  2.83 $\to$  
	2.88 &&  $2.91$\\
		\hline
		$\chi^2$ & & & & $12.4$ \\
		\hline		
	\end{tabular}
	\caption{Global best-fit of flavor observables within \textbf{MI}. Here CP violation is generated by a fixed phase $\omega$.} 
	\label{tab:fit1}
\end{table} 

One can perform a global fit to the flavor observables by defining the chi-square function
\begin{equation}
\chi^2=\sum (\mu_{exp}-\mu_{model})^2/\sigma^2_{exp},
\end{equation}
where the sum runs through the 19 measured physical parameters (note that the overall neutrino mass scale and the two Majorana phases are currently undetermined).
We make use of the MPT package \cite{Antusch:2005gp} to obtain the flavor observables from the mass matrices in Eq. (\ref{eq:massmat}).
Then we scan the 15 free parameters and find the values that minimize the $\chi^2$ function.
Neutrino oscillation parameters are taken from the global fit in Ref.~\cite{deSalas:2017kay}, while the rest of the observables are taken from the PDG~\cite{Tanabashi:2018oca}. For consistency of the fit, all quark and charged lepton masses are evolved to the same common scale, which we choose to be $M_Z$. The running of CKM and neutrino mixing parameters is negligible~\cite{Xing:2007fb,Antusch:2013jca}.
The results are shown in table \ref{tab:fit1}.
One sees from the fit that $\chi^2=12.4$. This indicates a relatively good global fit, with some tension in the description of quark CP violation, as seen from the table.
The origin for this is traced to the absence of a free parameter describing CP violation, as discussed above.

\subsection{Model Set-up II (MII)}
\label{sec:M2}
%%%%%%%%%%%%%%%%%%%%%%%%%%%%%%%%%%%%%%%%%%%%%%%%%%

We can now relax the assumption that all the Higgs VEVs are real, allowing them to be general complex numbers.
However, we keep the assumption that the Yukawa couplings are real. This reinstates 2 physical phases $ \phi_1^{\nu,d}-\phi_2^{\nu,d}$, now increasing to the number of free parameters to 17,
for a total of 22 flavor observables.
\begin{table}[ht]
	\centering
	\footnotesize
	\renewcommand{\arraystretch}{1.1}
	\begin{tabular}[t]{|lcr|}
		\hline
		Parameter &\qquad& Value \\ 
		\hline
		$y^e_1v_d/\mathrm{(10^{-1}GeV)}$ &\quad& $-1.020$ \\
		$y^e_2v_d/\mathrm{GeV}$ && $1.745$ \\
		\rule{0pt}{3ex}%
		$y^d_1v_d/(10^{-2}\mathrm{GeV})$ &&$-5.069$ \\
		$y^d_2v_d /\mathrm{GeV}$ && $2.869$ \\
		\rule{0pt}{3ex}%
		$y^\nu_{1} v_\nu/\sqrt{Y^N v_\sigma \mathrm{meV}}$ &&$-1.461$ \\
		$y^\nu_2v_\nu/\sqrt{Y^N v_\sigma \mathrm{meV}}$ && $7.647$ \\
		\rule{0pt}{3ex}%
		$y^u_{1} v_u/(10^{-1}\mathrm{GeV})$ &&$6.198$ \\
		$y^u_2v_u/(10^2\mathrm{GeV})$ && $1.712$ \\
		$y^u_3v_u/\mathrm{GeV}$ && $7.143$ \\
		\rule{0pt}{3ex}%
		$\epsilon^u_1/10^{-4}$ && $-6.926 $ \\
		$\epsilon^u_2/10^{-2}$ && $-5.058$ \\
		\rule{0pt}{3ex}%
		$\epsilon^d_1/10^{-3}$ && $2.812$ \\
		$\epsilon^d_2/10^{-3}$ && $5.863$ \\
		\rule{0pt}{3ex}%
		$\epsilon^\nu_1/10^{-1}$ && $-9.950$ \\
		$\epsilon^\nu_2/10^{-1}$ && $5.979$ \\
		\rule{0pt}{3ex}%
		$(\phi^d_1-\phi^d_2)/\pi$ && $-1.078$ \\
		$(\phi^\nu_1-\phi^\nu_2)/\pi$ && $1.093$ \\
		\hline
	
	\end{tabular}
	\hspace*{0.5cm}
	\begin{tabular}[t]{ |l |c|c c |c| }
		\hline
		\multirow{2}{*}{Observable}& \multicolumn{2}{c}{Data} & & \multirow{2}{*}{Model best fit}  \\
		\cline{2-4}
		& Central value & 1$\sigma$ range  &   & \\
		\hline
		$\theta_{12}^\ell$ $/^\circ$ & 34.44 & 33.46 $\to$ 35.67 && $34.65$  \\ 
		$\theta_{13}^\ell$ $/^\circ$ & 8.45 & 8.31 $\to$ 8.61  && $8.44$  \\  
		$\theta_{23}^\ell$ $/^\circ$ & 47.69 & 45.97 $\to$ 48.85  && $47.56$ \\ 
		$\delta^\ell$ $/^\circ$ & 237 & 210 $\to$ 275 && $198.3$  \\
		$m_e$ $/ \mathrm{MeV}$ & 0.489 &  0.489 $\to$ 0.489 && $0.489$ \\ 
		$m_\mu$ $/  \mathrm{GeV}$ & 0.102 & 0.102  $\to$ 0.102  &&  $0.102$ \\ 
		$m_\tau$ $/ \mathrm{GeV}$ &1.745 & 1.743 $\to$1.747 && $1.745$ \\ 
		$\Delta m_{21}^2 / (10^{-5} \, \mathrm{eV}^2 ) $ & 7.55  & 7.39 $\to$ 7.75 && $7.55$  \\
		$\Delta m_{31}^2 / (10^{-3} \, \mathrm{eV}^2) $ & 2.50  & 2.47 $\to$ 2.53 &&  $2.42$ \\
		$m_1$ $/\mathrm{meV}$  & & & & $24.31$ \\ 
		$m_2$ $/\mathrm{meV}$  & && & $25.81$ \\ 
		$m_3$ $/\mathrm{meV}$  & && & $55.60$ \\
		$ \phi_{12} $ $/^\circ$ & & && $252.5$  \\
		$ \phi_{13}$  $/^\circ$& & && $142.3$  \\    	
		$ \phi_{23}$  $/^\circ$& & && $51.5$  \\    	
		\hline
		$\theta_{12}^q$ $/^\circ$ &13.04 & 12.99 $\to$ 13.09 &&  $13.04$ \\	
		$\theta_{13}^q$ $/^\circ$ &0.20 & 0.19 $\to$ 0.22 && $0.20$  \\
		$\theta_{23}^q$ $/^\circ$ &2.38& 2.32 $\to$ 2.44 && $2.38$  \\	
		$\delta^q$ $/^\circ$ & 68.75 & 64.25 $\to$ 73.25  & & $69.25$\\
		$m_u$ $/ \mathrm{MeV}$ & 1.28 & 0.76$\to$ 1.81 && $1.29$  \\	
		$m_c$ $/ \mathrm{GeV}$ & 0.626 & 0.607 $\to$ 0.645 &&  $0.626$ \\	
		$m_t$ $/\mathrm{GeV}$  	  & 171.6& 170 $\to$ 173  && $171.6$ \\
		$m_d$ $/ \mathrm{MeV}$ & 2.74 & 2.35 $\to$ 3.15 &&  $2.51$ \\	
		$m_s$ $/ \mathrm{MeV}$ & 54 & 51 $\to$ 57 && $54$ \\
		$m_b$ $/ \mathrm{GeV}$	  & 2.85 & 2.76 $\to$ 2.94 &&  $2.87$\\
		\hline
		$\chi^2$ & & & & $1.6$ \\
		\hline
		
	\end{tabular}
        %\caption{Best fit for the model, setup II.}
        \caption{Global best-fit of flavor observables within \textbf{MI}. Here there are two free CP violation phases.} 
	\label{tab:fit2}
\end{table}

In this general setup, we loose predictivity for the physical CP violating phases $\delta^{l,q}$, leading to a drastic improvement of the global fit, achieving a minimum $\chi^2=1.6$ as seen in the table \ref{tab:fit2}.
Notice now the very good agreement of all of the observales, including the value of the CKM CP violation phase $\delta^{q}$.

\begin{figure}[h]
%%\centering
\includegraphics[height=7cm,width=0.6\textwidth]{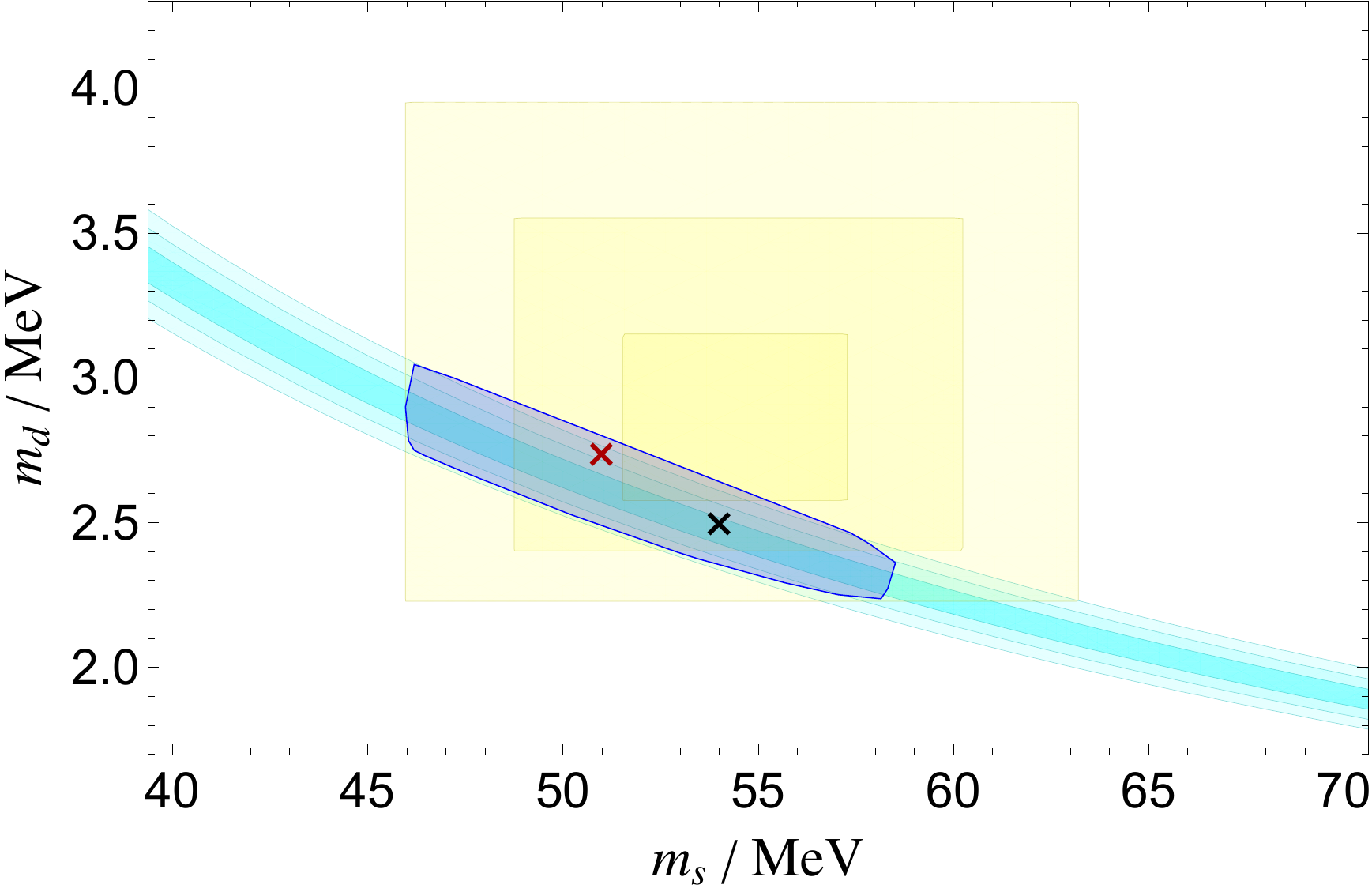}
\caption{ 
Prediction for the down-quark and strange-quark masses at the $M_Z$ scale.
  The cyan contours represent the 1,  2 and 3$\sigma$ allowed regions from the golden relation $m_\tau/\sqrt{m_\mu m_e}=m_b/\sqrt{m_s m_d}$.
  The yellow contours show the 1, 2 and 3$\sigma$ ranges of the measured quark masses at the $M_Z$ scale~\cite{Antusch:2013jca}.
  The blue region is the allowed parameter space consistent at 3$\sigma$ with the global flavour fit in Table \ref{tab:fit2}. The red (black) cross indicates the location of the best fit point for \textbf{MI} (\textbf{MII}).} 
\label{fig:golden}
\end{figure}

As mentioned above, a characteristic feature of our schemes is the golden quark-lepton mass relation given in Eq.~\ref{eq:golden}.
We now turn to study this prediction as obtained from our global fits of flavor observables within Models I and II.
In Fig. \ref{fig:golden} we use the golden quark-lepton mass relation in \textbf{MI} and \textbf{MII} to make predictions for the down and strange quark masses.
  Here the cyan bands stand for the 1, 2 and 3$\sigma$ regions compatible with the exact golden relation $m_\tau/\sqrt{m_\mu m_e}=m_b/\sqrt{m_s m_d}$ at the $M_Z$ scale, and the yellow contours are the 1, 2 and 3$\sigma$ regions for the quark mass parameters measured at the same scale.
To better appreciate the predictive power of our framework, we have varied randomly the parameters of \textbf{MII} around the best fit point in Table III and we have determined the shape of the parameter region consistent at 3$\sigma$ with all the 19 measured parameters of the model. This region is shown in purple in Fig. \ref{fig:golden}.
The corresponding contour for \textbf{MI} is not shown as it is very similar, given the fact that the golden relation is not very sensitive to the improvement of the CP violating phases in \textbf{MII}, compared to \textbf{MI}.
  However, one can see that the best fit point for scheme \textbf{MII}, indicated by the black cross, is now compatible at 1$\sigma$ with the exact golden formula.
%%%%%%%%%%%%%%%%%%%%%%%%%%%%%

%%%%%%%%%%%%%%%%%%%%%%%%%%%%%%%%%%%%
\section{Probing neutrino predictions}
\label{sec:neutrino-predictions}

In this section we present a close-up of the neutrino predictions of our orbifold compactification schemes, examining also the capability of future experiments to test them. 

\subsection{Neutrino oscillations at DUNE}
\label{sec:Dune}

%%%%%%%%%%%%%%%%%%%%%%%%%%%%%%%%%%%%%%%%%%%%%%%%%% the leptonic CP phase associated to within the three flavor paradigm

We start by quantifying the capability of the DUNE experiment to test the oscillation predictions resulting from the $A_4$ family symmetry, as realized from a six-dimensional spacetime after orbifold compactification.
Before presenting details about the simulated results, we first give a brief technical overview of the simulation details of DUNE, the proposed next generation superbeam neutrino oscillation experiment at Fermilab, USA~\cite{Acciarri:2015uup, Alion:2016uaj}.
The collaboration plans to use neutrinos from the Main Injector (NuMI) at Fermilab as a neutrino source.
In this experiment, the first detector will record particle interactions near the beam source, at Fermilab.
On the other hand, the neutrinos from Fermilab will travel a distance of 1300 km before reaching the far detector
situated at the underground laboratory of “Sanford Underground Research Facility (SURF)” in Lead, South Dakota.
The proposed far detector will use four 10~kton volume of liquid argon time-projection chambers (LArTPC).
The expected neutrino flux corresponding to 1.07 MW beam power gives $1.47\times 10^{21} $ protons on target (POT) per year for an 80 GeV proton beam energy.
We follow the same procedure as given in  \cite{Nath:2018fvw} for performing our numerical analysis of DUNE.
The \texttt{GLoBES}  package \cite{Huber:2004ka, Huber:2007ji} along with the auxiliary files as mentioned in ~\cite{Alion:2016uaj} has been utilized for the simulation.
We adopt 3.5 years running time in both neutrino and antineutrino modes, with a 40 kton total detector volume.
In the numerical analysis, we also take into account both the appearance and disappearance channels of neutrinos and antineutrinos.
In addition, both the signal and background normalization uncertainties for the appearance as well as disappearance channels have been taken into account in our analysis, as mentioned in the DUNE CDR \cite{Alion:2016uaj}.

Given that normal mass ordering (i.e., $m_1 < m_2 < m_3$) of neutrinos is currently preferred over the inverted one (i.e., $m_3 < m_1 < m_2$) at more than 3$ \sigma $
\cite{deSalas:2017kay}, we focus on the first scenario throughout this work.\\

In what follows, we examine the sensitivity regions of DUNE in the $(\sin^2\theta^{l}_{23}, \delta^{l} )$ for different seed points.
These are shown at $ 1\sigma $ (dark-orange), $ 2\sigma $ (orange), and $ 3\sigma $ (lighter-orange) confidence level, respectively.
%%%%
\begin{figure}[!t]
%\begin{center}
\includegraphics[height=7cm,width=10cm]{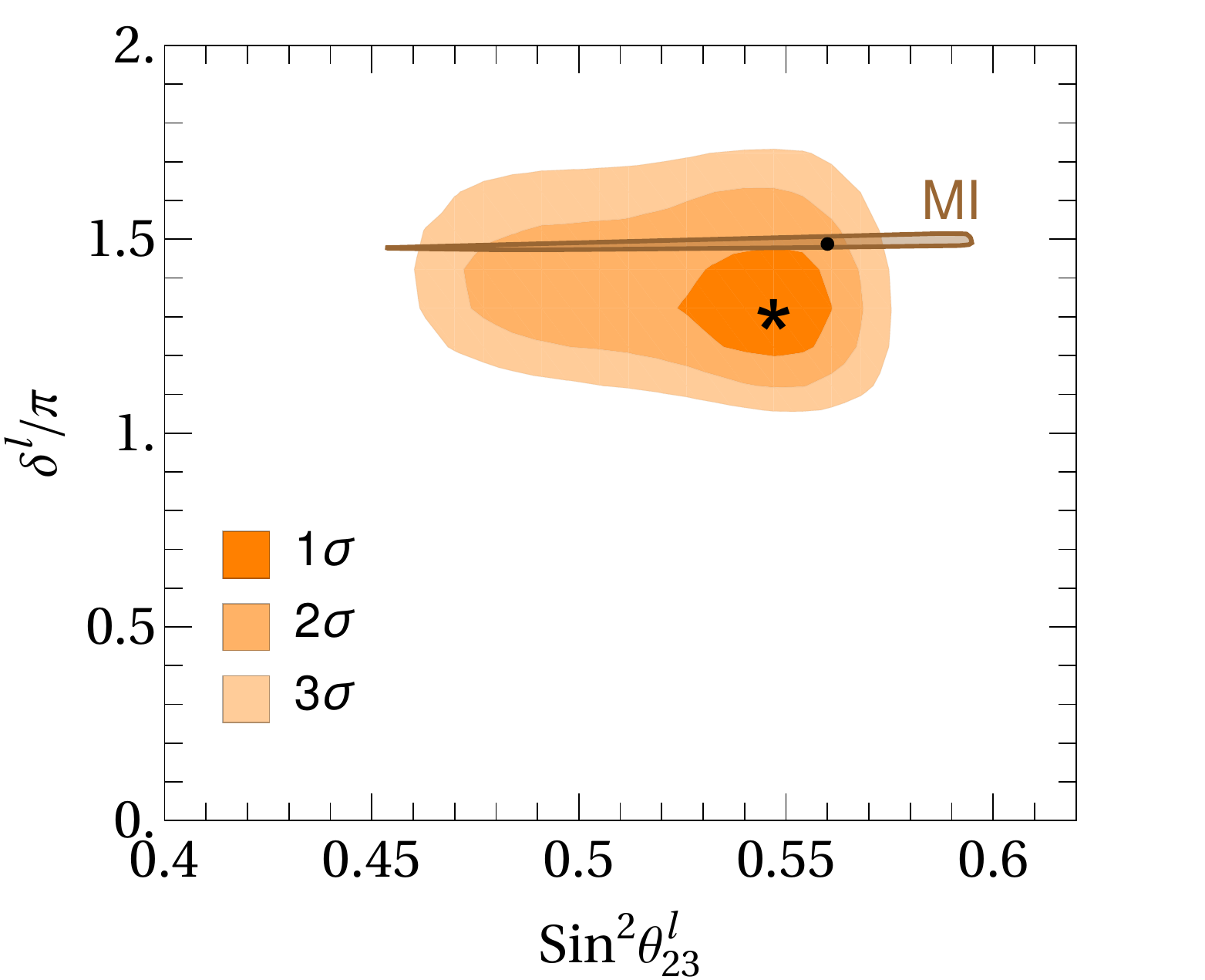}
\caption{\footnotesize  DUNE sensitivity region in the $(\sin^2\theta^{l}_{23}, \delta^{l} )$ plane. The `star-mark' represents the latest neutrino oscillation best-fit~\cite{deSalas:2017kay},
  while the `black-dot' is the predicted best-fit, as given in Table \ref{tab:fit1}.}
% \end{center}
\label{fig:DUNE-GF}
\end{figure}

%%%%%%%%%%%%%%%%%%%%%%%%%%%%%%%%%%%%%%%%%
\begin{figure}[!h]
%\begin{center}
\includegraphics[height=7cm,width=8cm]{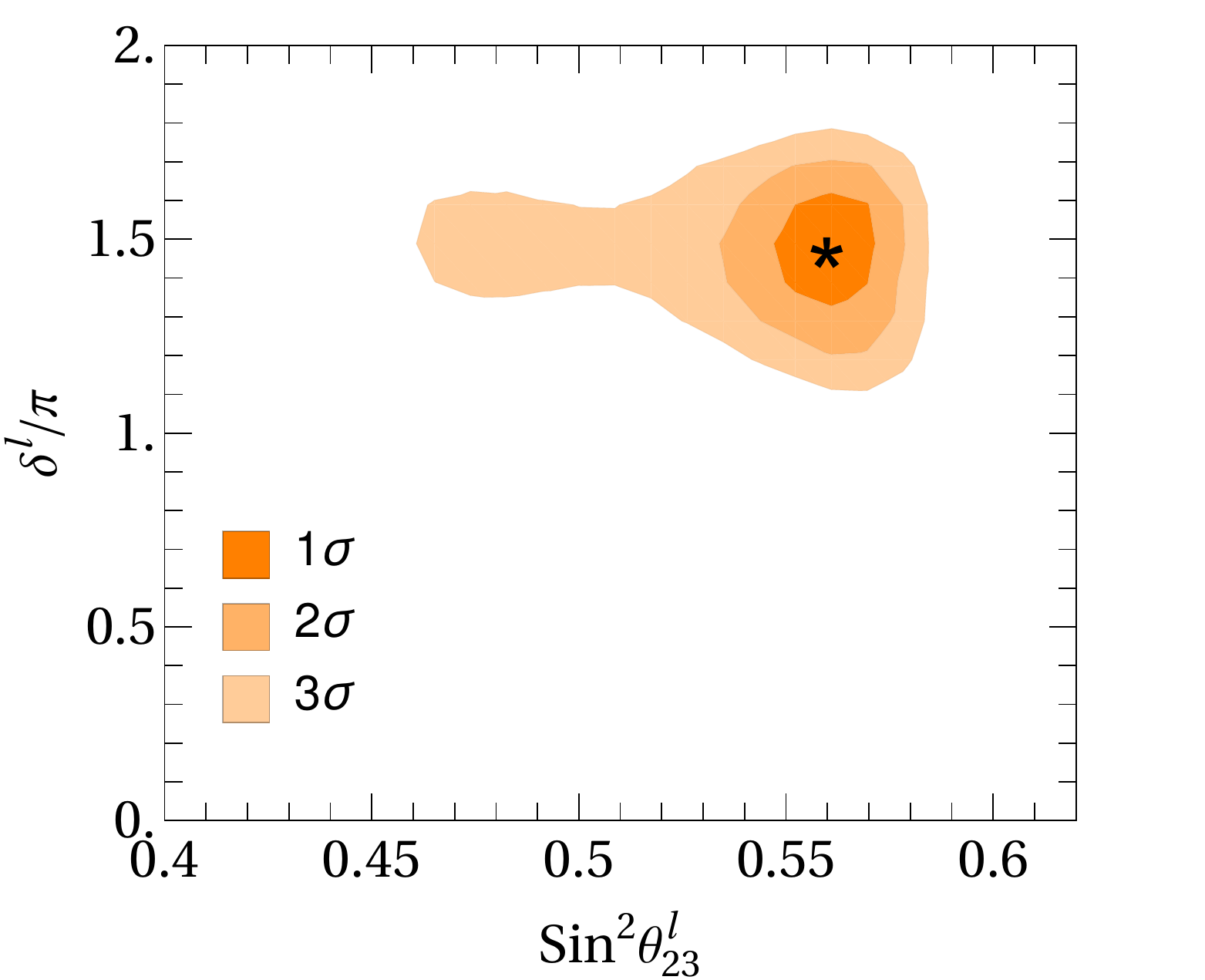}
\includegraphics[height=7cm,width=8cm]{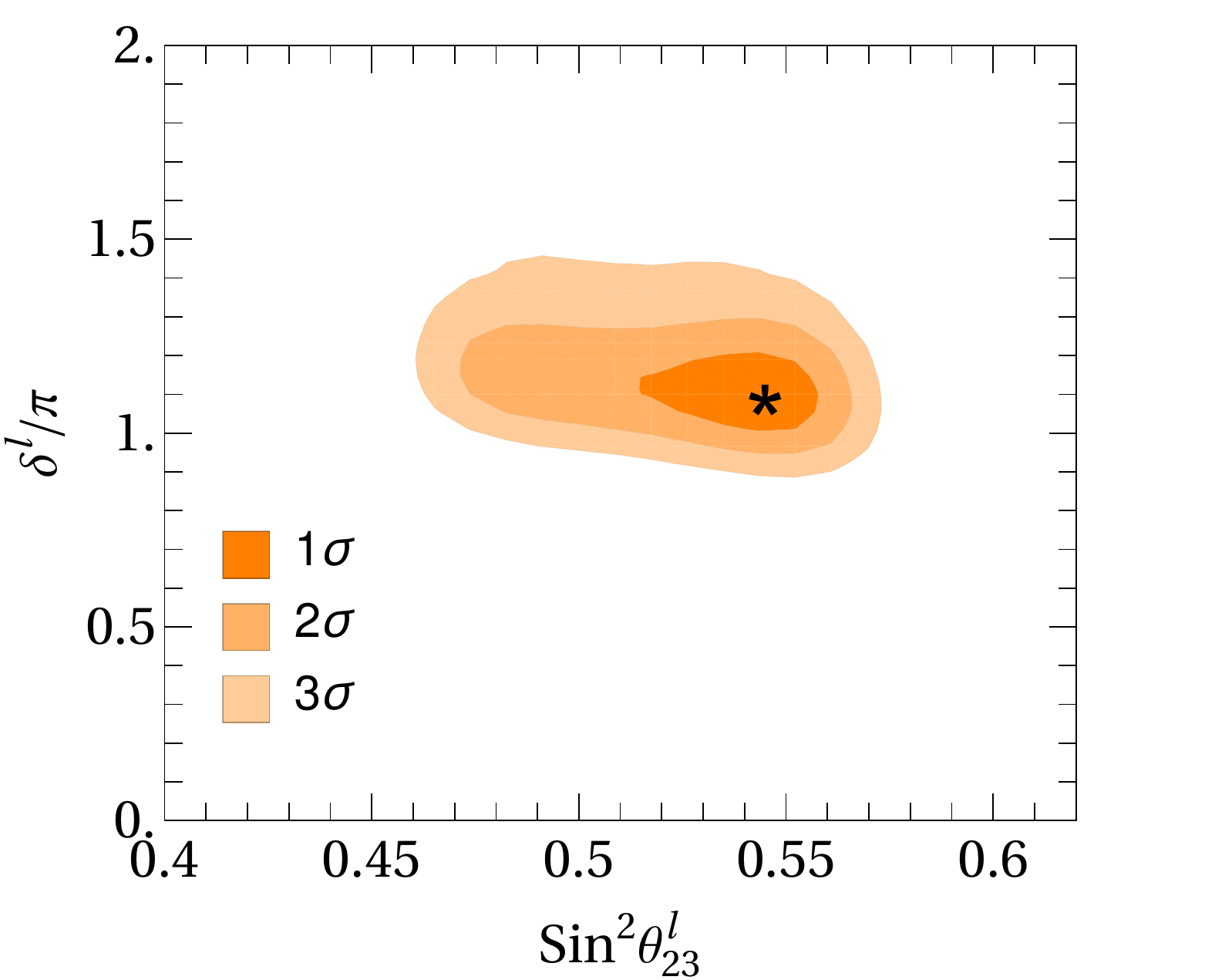}
\caption{\footnotesize DUNE $(\sin^2\theta_{23}, \delta^l )$ sensitivity regions in models {\bf MI} (left) and {\bf MII} (right),
  assuming the corresponding best-fit points obtained in setup {\bf MI} (left-panel) and {\bf MII} (right-panel), respectively, as indicated by the `star-marks'.
 }
% \end{center}
\label{fig:DUNE-Model}
\end{figure}
%%%%%%%%%%%%%%%%%%%%%%%%%%%%%%%%%%%%

In Fig. \ref{fig:DUNE-GF}, we show the expected 1,~2 and $3~\sigma$ DUNE sensitivity regions in the $(\sin^2\theta^{l}_{23}, \delta^{l} )$ plane. Here we assume model setup \textbf{MI} and take the latest neutrino oscillation best-fit point from~\cite{deSalas:2017kay} as benchmark, as indicated by the black `star-mark'. 
The corresponding \textbf{MI} theory predictions are indicated by the brown $ 3\sigma $ confidence level region, and its best-fit point, as given in Table~\ref{tab:fit1}, is shown by the black `dot'.
One sees that DUNE will be able to rule out predicted correlation between $\sin^2\theta^{l}_{23}, $ and $\delta^{l} $ for \textbf{MI} at $ 1\sigma $ C. L.
In contrast, we note that the predicted region in model \textbf{MII} covers the full DUNE sensitivity contours, so we do not show this plot in Fig. \ref{fig:DUNE-GF}.
  In other words, if the current best fit value of the oscillation parameters remains, DUNE will not be able to rule out the predictions for \textbf{MII} even at $ 1\sigma $ C. L.

We now change our seed points, adopting as benchmarks the $(\sin^2\theta^{l}_{23}, \delta^{l} )$ best-fit points predicted in each of the models described above.
The resulting DUNE sensitivity regions are given in Fig.~\ref{fig:DUNE-Model}.

One sees from the left-panel that, if the {\bf MI} predicted value of $(\sin^2\theta^{l}_{23}, \delta^{l} )$ is the true beanchmark value, then DUNE (after 3.5 running time in both neutrino and antineutrino modes) can rule out maximal value of $\theta^{l}_{23}$ i.e., $\sin^2\theta^{l}_{23} = \pi/2$ at $ 2\sigma $ confidence level.
On the other hand, by adopting the {\bf MII} predicted best-fit as the true seed value, we notice from the right-panel that DUNE can rule out maximal value of $\sin^2\theta^{l}_{23}$ at $ 1\sigma $, whereas it can rule out $\delta^{l}=3\pi/2$  at $3\sigma $ confidence level.

\subsection{Neutrinoless double-beta decay }
\label{sec:znbb}

\begin{figure}[!b]
%%\centering
\includegraphics[width=0.7\textwidth]{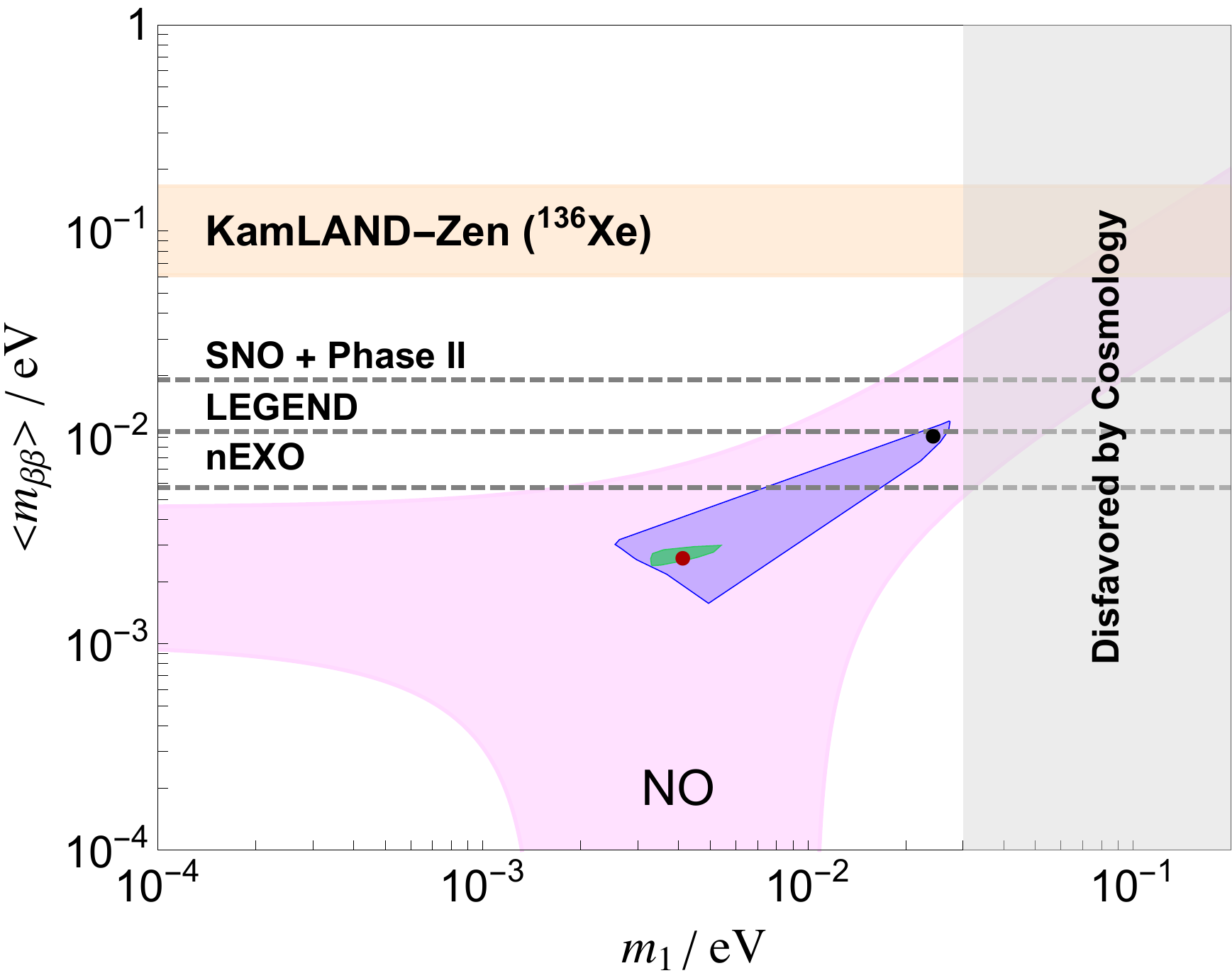}
\caption{ 
Effective Majorana neutrino mass parameter $\langle m_{\beta\beta}\rangle$ as a function of the lightest active neutrino mass $m_{1}$.
  %
%  The mass spectrum is normal ordered. The blue region is the generic one consistent with oscillations at 2$\sigma$.
%
Here green (blue) contour represents the predicted $\langle m_{\beta\beta}\rangle$ parameter space consistent at 3$\sigma$ with the global flavor fit for model setup MI (MII), and the best-fit value is shown by the red (black) dot.
  The current KamLAND-Zen limit is shown by the light-yellow band, and the projected sensitivities for future experiments are indicated in dashed horizontal lines, see text for details.} 
\label{fig:mee}
\end{figure}

We now turn to the predictions for neutrinoless double beta decay in Models {\bf MI} and {\bf MII} and confront them with experimental sensitivities~\cite{KamLAND-Zen:2016pfg,Alduino:2017ehq,Albert:2017owj,Agostini:2018tnm,Andringa:2015tza,Abgrall:2017syy,Albert:2017hjq,Azzolini:2020skx}.   This is shown in Fig. \ref{fig:mee}.
  The values for the effective mass $\vev{ m_{\beta\beta}}$ consistent at $3\sigma$ with the measured flavor observables (mainly neutrino oscillation parameters)
  obtained from the global fit are represented by the green contour for the case of the ``constrained'' model {\bf MI}, and by the blue one for {\bf MII}.
The theory-predicted regions are obtained by allowing the free parameters to vary randomly from the best fit point while simultaneously complying at 3$\sigma$ with all the measured observables of the global fit. One sees that the predicted region for {\bf MII} becomes wider, while the region for {\bf MI} remains quite small. This is due to the effect of the variation of the available free phases in {\bf MII} $ \phi_1^{\nu,d}-\phi_2^{\nu,d}$, which are directly related to the Majorana phases. In contrast, in {\bf MI} the only available CP violating phase is fixed, leading to sharply predicted \znbb decay amplitude which can not deviate much from its best fit value.

Interestingly enough, predictivity is not destroyed by the inclusion of those extra phases, and  \textbf{MII} still has upper and lower bounds for both the effective mass $\vev{ m_{\beta\beta}}$ and the lightest neutrino mass parameter. 
As a visual guide for the experimental searches of \znbb, in Fig. \ref{fig:mee} the horizontal yellow band indicates the current experimental limits from Kamland-Zen $(61 - 165\; \mathrm{meV})$ \cite{KamLAND-Zen:2016pfg}, while the dashed lines correspond to the most optimistic sensitivities projected for SNO + Phase II $(19 - 46 \; \mathrm{meV})$ \cite{Andringa:2015tza},  LEGEND $(10.7 - 22.8\; \mathrm{meV})$ \cite{Abgrall:2017syy}, and nEXO $(5.7 - 17.7\; \mathrm{meV})$ \cite{Albert:2017hjq}. One sees that the best fit point for MII, marked with a black point, becomes testable by the next generation of \znbb experiments LEGEND and nEXO. Finally, the vertical gray band represents the current sensitivity of cosmological data from the Planck collaboration \cite{Aghanim:2018eyx}.

\section{Conclusion }

We have investigated the implications of a recently proposed theory of fermion masses and mixings based on an $A_4$ family symmetry that arises from the compactification of a 6-dimensional orbifold.
We have analysed two variantions of the idea, a ``constrained'' one, in which CP violation is strictly predicted, and another where CP phases are free to vary.
We have quantified the predictions of these schemes for neutrino oscillations, neutrinoless double-beta decay and the golden quark-lepton mass formula.
We have found that the projected long baseline experiment DUNE can probe the model predictions concerning the maximality of the atmospheric mixing or the value of the CP phase in a meaningful way.
Likewise, the next generation of neutrinoless double-beta decay experiments, especially LEGEND and nEXO, could probe our model \textbf{MII} in a substantial region of parameters.

\acknowledgements 
\noindent

Work is supported by the Spanish grants SEV-2014-0398 and FPA2017-85216-P (AEI/FEDER, UE), PROMETEO/2018/165 (Generalitat Valenciana) and the Spanish Red Consolider MultiDark FPA2017-90566-REDC.  CAV-A is supported by the Mexican C\'atedras CONACYT project 749 and SNI 58928. NN is supported by the postdoctoral fellowship program DGAPA-UNAM, CONACYT CB-2017-2018/A1-S-13051 (M\'exico) and DGAPA-PAPIIT IN107118.

\bibliographystyle{utphys}
\bibliography{bibliography}
\end{document}